\documentclass[conference]{IEEEtran}
\IEEEoverridecommandlockouts
\usepackage{amsmath}
\usepackage{amsfonts}
\usepackage{graphicx}
\usepackage{epstopdf}
\usepackage{xcolor}
\usepackage{cite}
\usepackage{comment}
\usepackage{cases}
\usepackage{siunitx}
\usepackage{tikz}
\def\BibTeX{{\rm B\kern-.05em{\sc i\kern-.025em b}\kern-.08em
    T\kern-.1667em\lower.7ex\hbox{E}\kern-.125emX}}

\begin{document}

\title{ % Move title up to accommodate acceptance text
\text{\small Accepted in 2025 IEEE PES GENERAL MEETING, Austin, Texas (PES GM 2025)} \\ Detecting Unobservable Contingencies in Active Distribution Systems Using a Stochastic Hybrid Systems Approach 
\thanks{This work is supported in part by National Science Foundation under Grants DMS-2229109.\\
* Corresponding Author}}

%\author{\IEEEauthorblockN{Erfan Mehdipour Abadi, Hamid Varmazyari, Masoud H. Nazari~\IEEEmembership{Senior Member,~IEEE,}}
%\\\small Department of Electrical and Computer Engineering, Wayne State University, Detroit, MI, USA
%\\\small  erfan.abadi@wayne.edu, varmazyari.h@wayne.edu, masoud.nazari@wayne.edu}
%\date{}

\author{\IEEEauthorblockN{Erfan Mehdipour Abadi \\\small 
\textit{Electrical and Computer Engineering} \\
\textit{Wayne State University}\\
Detroit, USA \\
erfan.abadi@wayne.edu
\vspace{-2em}}
\and
\IEEEauthorblockN{Hamid Varmazyari \\\small \textit{Electrical and Computer Engineering} \\
\textit{Wayne State University}\\
Detroit, USA \\
varmazyari.h@wayne.edu
\vspace{-2em}}
\and
\IEEEauthorblockN{Masoud H. Nazari* \\\small \textit{Electrical and Computer Engineering} \\
\textit{Wayne State University}\\
Detroit, USA \\
masoud.nazari@wayne.edu}
\vspace{-2em}}

\maketitle

\begin{abstract}
    This paper introduces a distributed contingency detection algorithm for detecting unobservable contingencies in power distribution systems using stochastic hybrid system (SHS) models. We aim to tackle the challenge of limited measurement capabilities in distribution networks that restrict the ability to detect contingencies promptly.
    We incorporate the dynamics of distribution network connections, load feeders, PV, and battery energy storage system (BESS) hybrid resources into a fully correlated SHS model representing the distribution system as a randomly switching system between different structures during contingency occurrence. We show that jumps in the SHS model correspond to contingencies in the physical power grid.   
   We propose a probing approach based on magnitude-modulation inputs (MaMI) to make contingencies detectable. The effectiveness of the proposed approach is validated through simulations on a sample distribution system. 
\end{abstract}


%~~~~~~~~~~~~~~~~~~~~~~~~~~~~~~~~~~
\section{Introduction} \label{Sec1}
The integration of renewable energy resources and energy storage systems into distribution systems impacts their operational dynamics. 
The rising penetration of low-inertia, power electronic interfaced (PEI) resources makes these systems vulnerable against contingencies \cite{8700246}. 
Simultaneously, financial and environmental considerations drive the utilization of these resources close to their maximum capacity \cite{bulat2021enhanced}. These factors collectively present challenges to the reliability and resilience of distribution systems, demanding innovative solutions to address these issues \cite{ islam2024improving}.

Distribution systems often suffer from inadequate measurement capabilities, which hinder prompt detection of various contingencies \cite{ sim2024detection}. These contingencies, frequently go unnoticed due to their negligible initial impact on the system, occurrence behind measures, or lack of direct measurements. Such oversight can escalate into severe catastrophic outcomes \cite{che2019identification}.  Therefore, the challenge of prompt detection of undetectable contingencies must be tackled for the reliable operation of distribution systems \cite{poddubnyy2024semi}.

Traditionally, contingencies have been detected through signature analysis, which relies heavily on extensive training data \cite{ aminifar2021machine}. Additionally, there is a growing emphasis on developing advanced machine learning algorithms to detect various system anomalies and contingencies \cite{ehsani2022convolutional}. However, analyzing contingencies and assessing the system's reliability remains challenging due to the limited availability of analytical information about contingencies.

In \cite{CD1,yin2022joint}, we have shown that contingencies can be effectively modeled as stochastic jumps in system dynamics, represented as discrete events. Consequently, the distribution system dynamics encompass continuous physical variables (x), such as voltage and frequency, and discrete variables ($\alpha$) that characterize contingencies. This framework naturally forms a SHS model for distribution power systems.

In \cite{CD2, AMPS}, we applied the SHS model to transmission systems with synchronous generators. In this paper, we extend the SHS model to active distribution systems with PES resources, such as PV and BESS. To address this, we partition the system into segments, developing individual SHS models for each segment. Each model encompasses a PEI resource along with its associated measurement systems, enabling detection of all contingencies specific to that segment.

Undetectable contingencies in distribution systems typically affect only a portion of the system, leaving the rest unchanged. As a result, the system matrices under such contingencies, generally share common eigenvalues. Meanwhile, the initial condition of the system is undetermined due to the unknown exact occurrence time of the contingency. In \cite{CD2}, we demonstrated that due to the presence of common eigenvalues and unknown initial conditions, existing sensing networks are unable to detect unnoticed contingencies. Therefore, probing inputs should be introduced to distinguish these contingencies. \cite{yin2022joint} introduces two types of probing inputs that enhance the distinguishability of switching scenarios. This study compares various probing inputs and their effects on active distribution systems and proposes an effective design procedure for their implementation. The main contributions of this paper are presented below:
\begin{itemize}

   \item We extend the SHS modeling to active distribution systems with high penetration of PEI resources. This model represents contingencies as discrete events, creating a hybrid continuous-discrete framework for distribution systems. In this approach, the jumps in the SHS model correspond to the occurrence of contingencies. 
   %A fully correlated SHS model has been developed to detect unobservable contingencies in active distribution systems with a large penetration of PEI resources.  
    \item Detect unnoticed or unobservable contingencies in distribution systems, which may remain undetected during normal operations but become apparent during extreme events, such as storms, potentially leading to widespread outages.
    \item The effectiveness of the proposed probing input and contingency detection algorithm is demonstrated on a representative distribution system featuring PV and BESS resources, validating its reliability and robustness. The results demonstrate that the proposed method can effectively detect and distinguish various types of contingencies, even with limited sensing and monitoring capabilities.
\end{itemize}

The structure of this paper is organized as follows. Section \ref{Sec2} introduces the distributed SHS model developed for active distribution systems. Section \ref{Sec3} details the derivation of contingency detection based on the SHS model. The simulation results and performance evaluation are presented in Section \ref{Sec4}. Finally, the conclusions are summarized in Section \ref{Sec5}.

\section{Distributed SHS Model} \label{Sec2}

The SHS model is specifically developed for power systems in \cite{CD2}, within a centralized framework detailed as
\begin{equation}\label{eq_SHSLinearizedSS}
    \dot x = A(\alpha(t))x + B_1(\alpha(t)) u_1 + B_2(\alpha(t)) u_2,
\end{equation}
where $A(\alpha(t))$ represents system matrix, $B_1(\alpha(t))$ is the control input, and $B_2(\alpha(t))$ denotes disturbance matrix. They are known matrices that can be changed by $\alpha \in \mathcal{S}=\{1,2,\dots,m\}$ corresponding to various structural system configurations during contingencies. Also, $x$, $u_1$, and $u_2$ represent the system states, control inputs, and disturbances, respectively. Additionally, a contingency detection algorithm developed in \cite{CD2} identifies the active switching sequence of the power system within each $k'$-th interval $t\in [k\tau,k\tau + \tau_0)$, where $\tau_0\ll\tau$.
%~~~~~~~~~~~~~~~~~~~~~~~~~~~~~~~~~~
\subsection{Active Distribution System Segmentation} \label{Subsec1}
%\begin{comment}

\begin{figure}[t]
    \centering
    \vspace{-1.25em} % Reduce vertical space before the figure
    \includegraphics[width=1\linewidth, trim={0cm 4cm 2cm 1cm},clip]{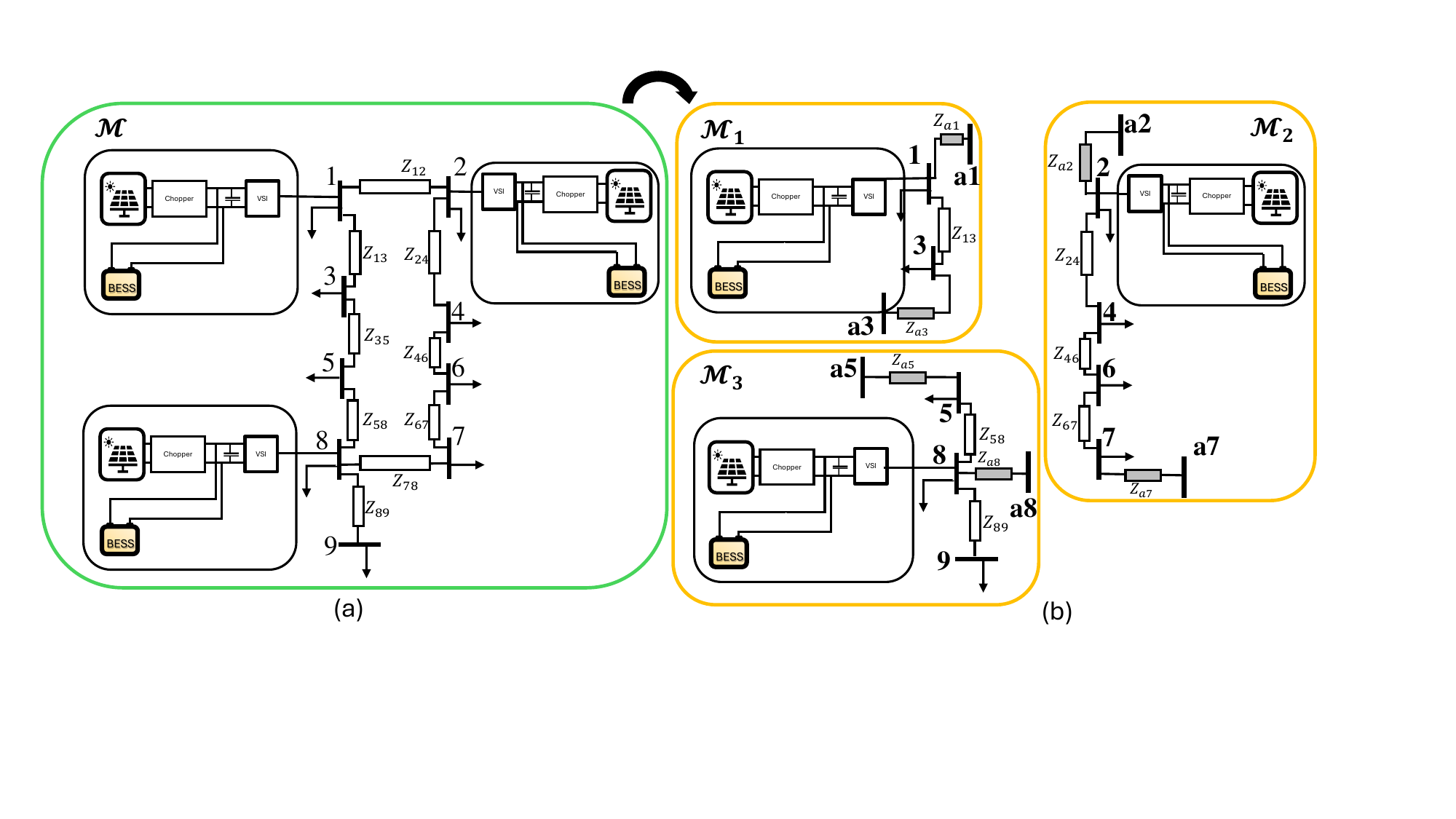}
    \caption{Architecture of the active distribution system.}
    \label{fig_islanded_microgrid}
    \vspace{-1.25em} % Reduce vertical space after the figure
\end{figure}

In this study, the distribution system is represented as a graph with two sets of nodes, $\mathcal{M} = (\Pi^{PV-B}, \Pi^L)$ for dynamic resources and load buses (see Fig. \ref{fig_islanded_microgrid} (a),) $\Pi^{PV-B}$ represents the set of all resource buses. Here, we consider a system consisting exclusively of PV and BESS hybrid resources (referred to as PV-B buses). However, altering the type of PEI source is expected to result in only minor modifications to the proposed approach.

All other buses $\Pi^L$ are considered as non-dynamic load buses. %Note that, non-dynamic buses are represented with a set of algebraic equations and could be written as a function of dynamic buses. We consider them aggregated to the dynamic load buses.
%Here we represent the load dynamics via a parallel RLC branch whose parameters could be adjusted according to any system characteristics \cite{sen2019simplified}.
The system's structure which represents graph edges is represented via the $\textbf{Z}$ where $Z_{ij} = R_{ij} + j X_{ij}$ represents the impedance between bus $i$ and $j$. 
To facilitate distributed monitoring of the system, we implement the system segmentation technique outlined in \cite{engelmann2018toward}. This approach employs auxiliary buses to partition the system into smaller segments,
$\mathcal{M}_i$, as illustrated in Fig. \ref{fig_islanded_microgrid}(b). Each segment $\mathcal{M}_i=(\Pi_i^{PV-B},\Pi_i^{L})$ contains source and resource buses 
where $\Pi_i^L \cap \Pi_j^L = \emptyset $ for all different segments $i \neq j$. 
$\mathcal{M}_i$ includes only a single PV-B bus defined by the singleton $\Pi_i^{PV-B}$. The structure of each segment is presented via $\textbf{Z}_i$.  
Load buses are assigned to these segments based on criteria such as geographical proximity. Note that, assigning a load bus to a particular $\mathcal{M}_i$ does not mean its demand will be exclusively supported by that segment; rather, indicates that the bus will be monitored by the segment's SHS model.

We define $\mathcal{N}_i$ as the set of neighboring buses for each $\mathcal{M}_i$. For instance, as shown in Fig. \ref{fig_islanded_microgrid}(b), the neighbor sets are $\mathcal{N}_1 = \{2,5\}$, $\mathcal{N}_2 = \{1,8\}$, and $\mathcal{N}_3 = \{3,7\}$. To facilitate separation between segments, an auxiliary bus, denoted as $a_i$, is introduced. When two auxiliary buses, $a_i$ and $a_j$, are assigned to separate the connection between buses $i$ and $j$, the impedance connected to these auxiliary buses is defined as $Z_{a_j}=Z_{a_i}=\frac{Z_{ij}}{2}$ within both $\textbf{Z}_i$ and  $\textbf{Z}_j$, and the auxiliary buses are coupled, meaning that $V_{a_i}=V_{a_j}$.

%~~~~~~~~~~~~~~~~~~~~~~~~~~~~~~~~~~

\subsection{SHS Model of Segment $\mathcal{M}_i$}

\begin{figure}[t]
    \centering
    \vspace{0em} % Reduce vertical space before the figure
    \includegraphics[width=0.9\linewidth, trim={6cm 4cm 7cm 4cm},clip]{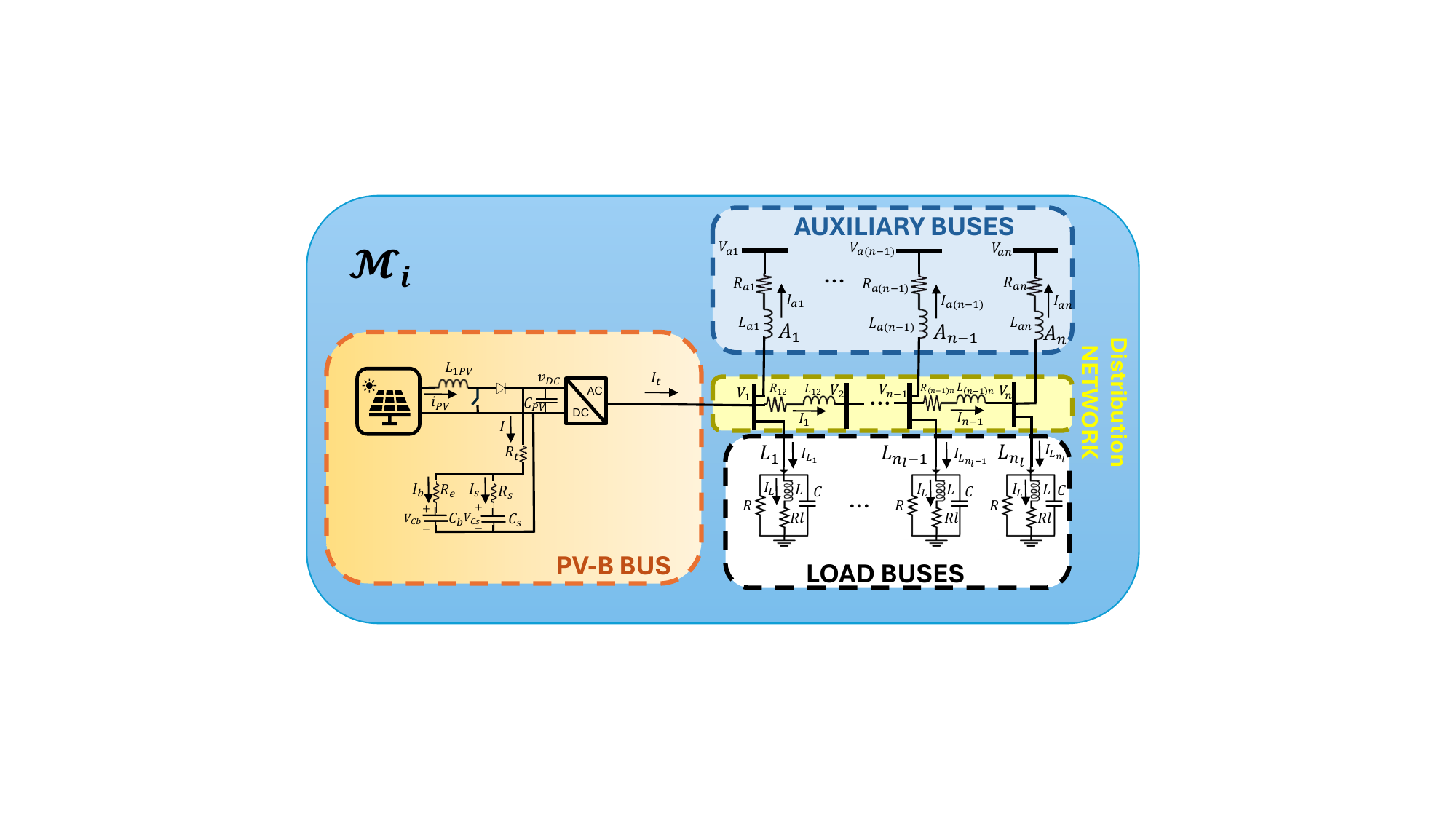} % Adjust the width as needed
    \vspace{-1em} % Reduce vertical space after the figure
    \caption{A generic representation of each segment of the distribution system.}
    \label{fig:SeperatedBusSet}
\end{figure}

The state space of $\mathcal{M}_i$ must be derived for all switching scenarios $\alpha \in \mathcal{S}$, separately. Note that, this procedure is done offline and once for each segment which follows the structure of Fig. \ref{fig:SeperatedBusSet}. In this model, the PV-B bus is connected to Bus 1. Also, if any bus $k \in \Pi^i$ is connected to a neighboring bus $z \in \mathcal{N}_i$, an auxiliary bus $a_k$ is connected to $k$ accordingly.

Consequently, to consider the dynamics of various buses, the correlated model of $\mathcal{M}_i$ is developed, and its state space model for each scenario $\alpha$ would be
\begin{equation} \label{MGStateSpace}
    \dot {x} = \begin{bmatrix}
        A_{11} & A_{12} & A_{13} & A_{14} \\
        A_{21} & A_{22} & A_{23} & A_{24} \\
        A_{31} & A_{32} & A_{33} & A_{34} \\
        A_{41} & A_{42} & A_{43} & A_{44} \\
    \end{bmatrix} x + \begin{bmatrix}
        B_{11} \\
        B_{12} \\
        B_{13} \\
        B_{14}
    \end{bmatrix} u_1+\begin{bmatrix}
        B_{21} \\
        B_{22} \\
        B_{23} \\
        B_{24}
    \end{bmatrix} u_2,   
\end{equation}
where $x:= [{x}_{PV-B} \text{ }  {x}_{Net} \text{ } {x}_{Aux} \text{ }     {x}_{Load}]$. More specific dynamics of each part of $\mathcal{M}_i$ are presented as follows.

\subsubsection{\textbf{PV-B Bus}}
 We consider a PV system connected to the distribution network through a DC-DC chopper and a DC-AC inverter and a DC link is placed between the inverters. The BESS model is presented via a generalized model \cite{bhangu2005nonlinear}. 
The state space of PV and BESS are developed in \cite{sen2019simplified} independently. However, their interconnected model must be developed by connecting the terminal voltage of the BESS to the DC link.

Since the disturbances of the PV-B bus terminal voltages, are equivalent to the dynamics of loads, PV-B disturbances are included within the $\mathcal{M}_i$ dynamics by the correlation between $x_{PV-B}$ and $x_{Load}$ with $A_{14}$ matrix. Any $\alpha \in\mathcal{S}$ that represents contingencies of PV-B affects $A_{11}$, and failures on Bus 1 affects $A_{14}$.     
Hence, the first row of \eqref{MGStateSpace} would be
\begin{equation}
    \dot{x}_{PV-B} = A(\alpha)_{11} x_{PV-B} + A(\alpha)_{14} x_{Load} + B_{11} u_1,
\end{equation}
where $x_{PV-B} = [i_{pv}\, v_{dc}\, i_{t,q}\, i_{t,d}\, V_{C_s}\, V_{C_b}]^T$ are the states of PV-B sources. $u_1= [d\, \delta\, m_a]^T$ is the control inputs including chopper duty cycle \( d \), phase angle \( \delta \), and inverter modulation index \( m_a \). $x_{Load}$ is explained in the following part 4. $A_{11}$, $A_{14}$, and $B_{11}$ are derived inspired by \cite{sen2019simplified} and $A_{12}=A_{13}=B_{21}=\textbf{0}$ due to lack of correlation between PV-B states with the dynamics of distribution network states and auxiliary buses.

%=====================================
\subsubsection{\textbf{Distribution Network}}
The dynamics of the distribution network are derived using KVL equations \cite{yu2015analysis}. This dynamic varies based on contingencies, $\alpha \in \mathcal{S}$, that affects the network connections for which KVL is applied. Here, $x_{Net}=[I_{1,q} I_{1,d} \cdots I_{(n-1),q} I_{(n-1),d}]^T$ represents the current flow dynamics of the distribution network. The correlation between the dq-axis currents and bus voltage dynamics is captured by matrices $A_{22}$ and $A_{24}$, respectively. 
Hence, we can formulate dynamics of $x_{Net}$ as
\begin{equation}
    \dot{x}_{Net} =  A(\alpha)_{22} x_{Net} + A(\alpha)_{24} x_{Load}.
\end{equation}

%===========================
\subsubsection{\textbf{Auxiliary Buses}}
The SHS model for auxiliary buses is derived by KVL equations with $x_{Aux}=[I_{a1,q}\,I_{a1,d} \cdots I_{a(n-1),q}\,I_{a(n-1),d}]^T$. Since the control agent in $\mathcal{M}_i$ lacks access to the dynamics of auxiliary buses affected by neighboring segments, we integrate the effect of auxiliary bus voltage variations as the disturbance of system  $u_2=[V_{a1,q}\,V_{a1,d} \cdots V_{an,q}\,V_{an,d}]^T$ which is transferred from the neighboring control agents to $\mathcal{M}_i$.

%Note that, the changes in the neighboring segments of $\mathcal{M}_i$ are integrated as disturbances to this segment by considering  $u_2=[V_{a1,q}\,V_{a1,d} \cdots V_{an,q}\,V_{an,d}]^T$ are transferred from the neighboring control agents to $\mathcal{M}_i$.

Consequently, we can formulate the dynamics of $x_{Aux}$ as
\begin{equation}
    \dot{x}_{Aux} = A_{33}(\alpha)x_{Aux}+  A_{34}(\alpha)x_{Load} + B_{23} u_2,
\end{equation}
where $A_{33}$ has similar structure to $A_{22}$ with replacing the distribution network impedance, $Z_{ij}$, values to auxiliary impedance, $Z_{ai}$, values and the effects of $A_{24}$ are represented by a combination of $A_{34}$ and $B_{23}$.

\subsubsection{\textbf{Load Buses}}
The dynamics of the loads within the $\mathcal{M}_i$ segment are calculated using KVL and KCL equations on each load under contingencies $\alpha \in \mathcal{S}$, where  $x_{Load}=[V_{1,q}\, V_{1,d}\, I_{L_{L1,q}}\,I_{L_{L1,d}} \cdots V_{n,q}\, V_{n,d}\,  I_{L_{Ln,q}} \, I_{L_{Ln,d}}]^T$. For deriving the KCL equations, the currents of that particular bus are equivalent to the states of $x_{PV-B}$, $x_{Net}$, and $x_{Aux}$. Thus, the last row of \eqref{MGStateSpace} is given by
\begin{align} \notag
    \dot{x}_{Load} =&  A(\alpha)_{41} x_{PV-B} + A(\alpha)_{42} x_{Net}+A(\alpha)_{43} x_{Aux}\\ &+ A(\alpha)_{44} x_{Load},
\end{align} 
where $A(\alpha)$'s are driven based on system physics similar to previous parts. 

The measurement infrastructure for each segment of $\mathcal{M}_i$ is represented by 
\begin{equation}\label{eq_SHSLinearizedY}
    y = C(\alpha(t))x + D_2(\alpha(t)) u_2, 
\end{equation}
where $C(\alpha(t))$ and $D_2(\alpha(t))$ denote observer matrices for system states and auxiliary bus voltages.
By defining ${C}({\alpha})$ to measure the DC link and terminal current of PV-B, and ${D}_{2}({\alpha})$ as the indicator for auxiliary bus voltage, we gain the observability of $\mathcal{M}_i$.

%==========================
\section{contingency detection based on SHS framework} \label{Sec3}
In \cite{CD2}, contingency detection for transmission systems is based on the SHS model, where contingencies are treated as random switches between a finite set of system configurations. Note that, the system's structure is assumed to be fixed during any time intervals $ t \in(k \tau, (k+1)\tau)$, and the system's operation is presented as a random switching sequence which is detected within a short segment of the time interval $ t \in(k \tau, k \tau + \tau_0)$. Thus, we can split each time interval into contingency detection and reliable operation of the system.  

The contingency detection algorithm utilizes the real-time system model introduced in Section \ref{Sec2} for estimation of unknown initial states and deriving the expected output of the system for different scenarios. Next, by comparing the system measurement with the expected outputs of different scenarios and choosing the scenario with minimum estimation error, the contingency is detected. Refer to \cite{CD2} for more details of this algorithm. 

To make the system's output distinguishable, two types of probing inputs are introduced \cite{yin2022joint}: 1. Mode-Modulated Input (MoMI) and 2. Magnitude-Modulated Input (MaMI). The effects of these inputs are discussed in the following.

\subsection{MoMI vs. MaMI Probing}
The MoMI is specifically designed to exhibit a mode that is distinct from the system’s modes across various contingencies, eliciting unique responses from the system for each contingency. Although, \cite{CD2} demonstrates the use of MoMI in detecting contingencies, its implementation in distribution systems presents substantial challenges.

The MoMI uses signals with variable frequencies applied over short time intervals, $\tau_0$, which are repeated frequently every $\tau$ seconds. This frequent alteration can lead to system inefficiencies, such as device heating and accelerated aging. Moreover, as distribution systems are complex systems containing diverse equipment, the potential for a signal’s frequency to coincide with the natural frequency of any component could lead to resonance and unforeseen complications. 
Additionally, the slow dynamics of the sources, relative to the frequency changes in MoMI, complicate the immediate detection of contingencies within the typically short detection windows.

In contrast, MaMI aims to use changes in the magnitude of the input to overcome the effects of initial condition uncertainties on the system output, making it easier to implement in distribution systems by alteration in the switching of PEI. This study focuses on designing an appropriate MaMI that is not only effective but also practical for real-world distribution system applications.

\subsection{MaMI Design}
The conditions of using MaMI for distinguishing the switching scenarios are described theoretically in \cite{yin2022joint}, where the probing input is in the form of
\begin{equation} \label{eqMAIM}
    u(t) = Ru_1(t),
\end{equation}
where $U_1(s)$ = $\mathcal{L} \{ u_1(t) \}$ = $\frac{b(s)}{a(s)}$, $a(s)$ and $b(s)$ are polynomials of $s$ with real coefficients, and the order of ${b(s)}$ is strictly less than ${a(s)}$. We consider the step function as $u_1(t)$ that satisfies this condition. For distinguishing different scenarios, the effect of input $u(t)$ must overcome the unknown initial condition $x(0)$. To achieve this goal, $R$ must be greater than $R_0$, where $R_0$ is defined as 
\begin{equation} \label{MaMIFormula}
    R_{0} = \frac{2\mu_{0}\mu_{1}}{\delta_{min}},
\end{equation}
Here, $\mu_{0}$ is the upper bound for the norm of the initial condition of the system states.
 \begin{equation}
     ||x(0)||\leq \mu_{0}.
\end{equation}

Due to the physical limitation, the infimum of $\mu_{0}$ must be considered for designing the probing input. The $l$-inf norm ($||.||_{inf}=\max\{.\}$) serves as the design criterion where, the maximum perturbation of the system states must be chosen as \(\mu_{0}\). Assuming the advanced control mechanisms in PV-B and system voltage, we consider \(\mu_{0}\) to represent a small perturbation—approximately 2\% of the maximum line or load current in the steady state derived from power flow analysis. 

According to \cite{yin2022joint}, $\mu_{1}$ is defined  as 
\begin{equation}
     \max_{\alpha\in S}\max_{t\in[0, \tau_{0}]}||C(\alpha)\exp^{A(\alpha)t}||\leq \mu_{1}.
\end{equation}

We assume that unobservable contingencies are minor enough not to cause system instability, as any instability would propagate and be detectable by the existing sensor infrastructure. Thus, $A(\alpha)$ consists of negative eigenvalues, and $\exp^{A(\alpha)t}$ is monotonically decreasing for $t\in[0, \tau_{0}]$. Consequently, we have

\begin{equation}
    \mu_{1} =  \max_{\alpha\in S}||C(\alpha)||,
\end{equation}
at $t=0$.

As the system outputs for different contingency scenarios are compared for contingency detection, the min-max criteria is chosen for  $\delta_{min}$ as
\begin{equation}
    \delta_{min} = \min_{\substack{ i,j\in S\\ i \neq j}}\max_{t\in[0, \tau_{0}]} |y^{0}_{i}(t)- y^{0}_{j}(t)|.
\end{equation}
This formulation ensures that $\delta_{min}$  represents the smallest maximum deviation between output trajectories of any two distinct contingency scenarios, which helps in distinguishing them effectively by maximizing the separability of the closest scenarios in terms of their system responses.
%~~~~~~~~~~~~~~~~~~~~~~~~~~~~~~~~~~
\section{Simulation} \label{Sec4}
\begin{figure}[t]
    \centering
    \vspace{-0em} % Reduce vertical space before the figure
    \includegraphics[width=0.8\linewidth, trim={6cm 6cm 9cm 1cm},clip]{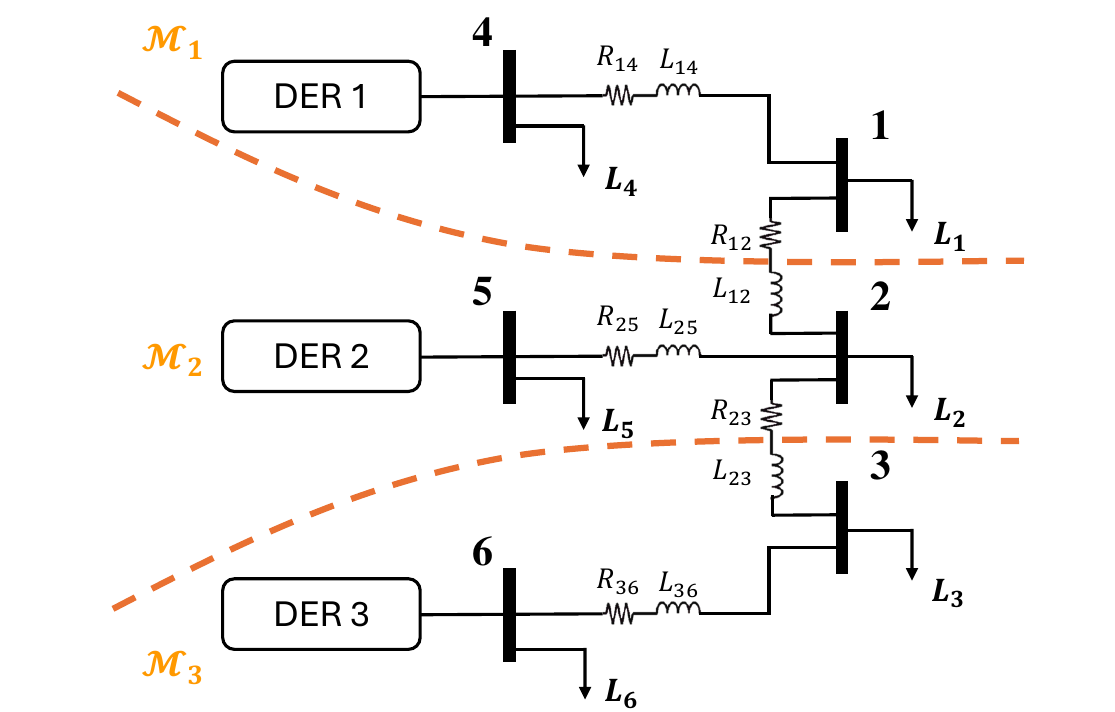} % Adjust the width as needed
    \vspace{-1em} % Reduce vertical space after the figure
    \caption{The PV-B Integrated Distribution System.}
    \label{fig:SimulationFig}
\end{figure}

To evaluate the functionality of SHS-based contingency detection, we simulate a sample distribution system setup similar to that described in \cite{vazquez2018fully}, as illustrated in Fig.~\ref{fig:SimulationFig}. 
It is an equivalent model of a distribution system with six dynamic nodes. The total demand of the system is 237 MVA. The system has 3 identical PV-B systems across buses 4, 5, and 6 with 100 MVA capacity. Details of the PV-B units and distribution network parameters are presented in Table~\ref{tab:DER Params}. The load parameters are specified in Table~\ref{tab:load Params}.
This simulation is developed within the MATLAB environment, utilizing the SHS model detailed in Section~\ref{Sec3}. The system is segmented into three parts: $\mathcal{M}_1$, $\mathcal{M}_2$, and $\mathcal{M}_3$, with each PV-B modeled as a $6^{th}$ order system, each distribution network line and auxiliary connection as $2^{nd}$ order, and each load bus as a $4^{th}$ order system. Consequently, the system orders for $\mathcal{M}_1$, $\mathcal{M}_2$, and $\mathcal{M}_3$ are 18, 20, and 18, respectively.

\begin{table}[t]
\vspace{-1em}
    \centering
        \caption{PV-B bus AND Line Parameters}
        \vspace{-0em}   
    \label{tab:DER Params}
\begin{tabular}{c c | c c} 
 \hline
 Parameters & Values & Parameters & Values \\ [0.5ex] 
 \hline\hline  
$R_{PV}$ & $-2.3\Omega$ &$R_{12}$ & $1.26$ \si{\ohm} \\ 
 $I_{PV}$ & $550A$  & $L_{12}$ & $0.636$ \si{\milli\henry} \\
  $L_{{1PV}}$ & $2mH$ & $R_{14}$ & $1$ \si{\ohm} \\
  $C_{PV}$ & $10mF$ & $L_{14}$ &  $0.7$ \si{\milli\henry} \\
  $R_{{2PV}}$ & $5.25m\Omega$  & $R_{23}$ &  $0.7$ \si{\ohm}\\
  $L_{{2PV}}$ & $1.8mH$    & $L_{23}$ & $3.686$ \si{\milli\henry}  \\
  $R_s$ &  $3.75m\Omega$   & $R_{25}$ & $0.8$ \si{\ohm} \\
  $R_e$ &  $3.75m\Omega$   &  $L_{25}$ &   $0.5$ \si{\milli\henry} \\
  $R_t$ &  $2.745m\Omega$   & $R_{36}$ &  $0.5$ \si{\ohm} \\
  $C_s$ &  $7586.5F$   & $L_{36}$ & $0.1$ \si{\milli\henry}  \\
  $C_b$ &  $7586.5F$   & &   \\ [1ex] 
 \hline
\end{tabular}
\end{table}
\begin{table}[t]
\vspace{-1em}
    \centering
        \caption{Load Parameters}
    \vspace{-1em}    
    \label{tab:load Params}
        \begin{tabular}{|c|c|c|c|c|c|c|}
            \hline
            Parameter &  $L_1$ & $L_2$ & $L_3$ & $L_4$ & $L_5$ & $L_6$ \\ \hline \hline
            $P (kW)$ & 45  & 22.5 & 66 & 25 & 12.5 & 36.67 \\ \hline
            $Q(kVAR)$  & 21.74  &10.89 &40.9 &12.09 &6.05 &22.72\\ \hline
            Power Factor   & 0.9  & 0.9 & 0.85 & 0.9  & 0.9 & 0.85\\ \hline
%            Inductor quality Factor & 130  &130 &120\\ \hline
            $R(\Omega)$ &  0.423 & 0.21 & 0.29 & 0.235 & 0.12&0.16\\ \hline
            $L(mH)$ &  180 & 90 & 270 & 100 &50 &150\\ \hline
            $Rl(\Omega)$ & 0.43  & 0.22 & 0.71 & 0.24 & 0.12 & 0.39\\ \hline
            $C(mF)$ &  1.29 & 0.65 & 2.45 & 0.72 & 0.36 & 1.36\\ \hline
        \end{tabular}%
        \vspace{-0em}
\end{table}

Due to space constraints, our analysis concentrates on the segment $\mathcal{M}_1$ for distributed monitoring under normal operations and three distinct contingency scenarios are considered as follows: 1. short circuit on Line 14, 2. single line outage on Line 14, and 3. line disconnection on Line 14. Thus, $\mathcal{S}=\{\alpha_0,\alpha_1,\alpha_2,\alpha_3\}$ where $\alpha_0$ indicates the normal operation; $\alpha_1$, $\alpha_2$, and $\alpha_3$ represent the contingency scenarios.

\begin{figure}[t]
    \vspace{-0em} % Reduce vertical space before the figure
    \centering
    \includegraphics[width=0.8\linewidth]{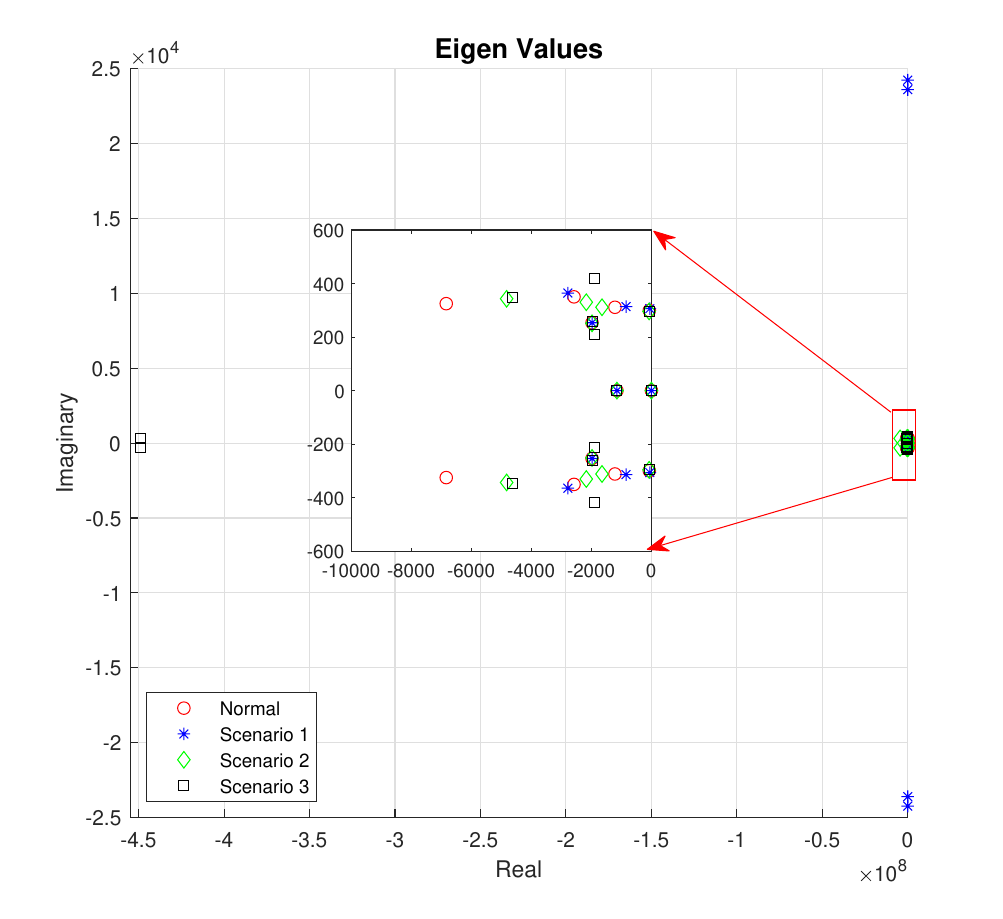}
    \vspace{-0.25em}
    \caption{Eigenvalue analysis under line 1-4 variations for normal, short circuit, single line outage, and line disconnection.}
    \vspace{-0.25em}
    \label{Fig_Eigenvalues}
\end{figure}

The eigenvalues of $\mathcal{M}_1$ under the four conditions are shown in Fig. \ref{Fig_Eigenvalues}. As shown in the figure, all eigenvalues lie on the left side of the complex plane, indicating stable operation of the system. However, they exhibit different dynamic responses due to changes in their eigenvalues. For instance, the damping factor of $\alpha_3$ is greater than the others, as shown by a more negative real part. $\alpha_1$ demonstrates a larger oscillatory behavior due to a larger imaginary part. Despite these variations in eigenvalues, some eigenvalues are common across the scenarios, necessitating use of probing input to detect contingencies.
Next, the probing input is designed and applied to the control input $\delta$ with parameters selected as follows: $\tau_0 = 10 ms$ for the detection period and $t_s = 1\mu s$ for the measurement sampling rate. 
Accordingly, the MaMI parameters are derived as follows:
$\mu_0$ is derived based on power flow analysis under normal conditions where $\mu_0 = 0.02 * \max(I_{ij}) =$ 5.63, and
$\mu_1 = ||C|| $ is fixed for $\alpha \in \mathcal{S}$. Since $C$ represents the direct measurement of certain states, $||C|| = 1$. 
Also, $\delta_{\min}$ = 112.15, which is derived based on comparing the system output deviations for 6 different combinations. Hence MaMI is defined as \eqref{eqMAIM} with $R=0.101$ and $u_1(t)= [0 \text{ } 1 \text{ } 0]^T$ which represents step function for the input $\delta$. 

\begin{figure}[t]
    \vspace{-1em} % Reduce vertical space before the figure
    \centering
    \includegraphics[width=0.8\linewidth, trim={0cm 1cm 1cm 1.5cm},clip]{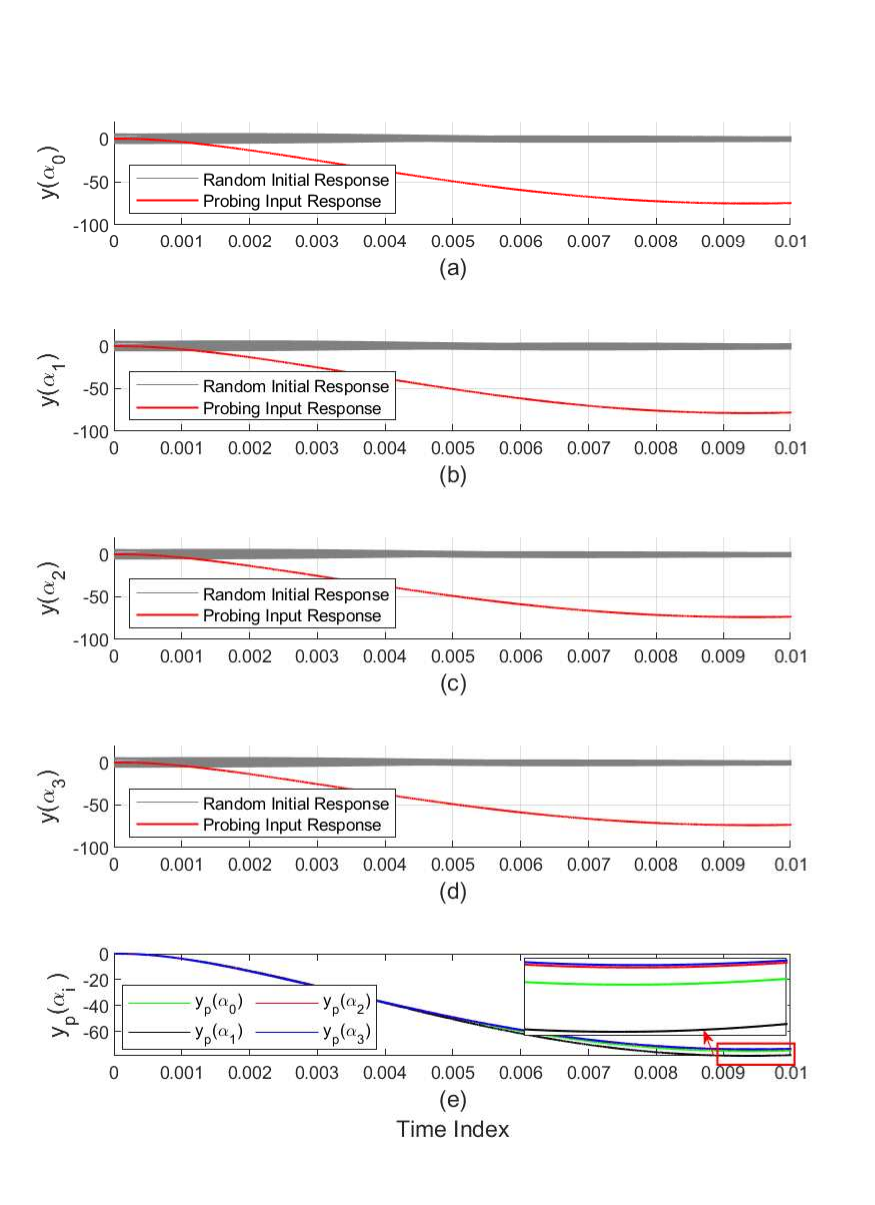}
    \vspace{-1em} % Reduce vertical space after the figure
    \caption{System responses for probing input ($y_p$) and random initial conditions for $\alpha_0$ to $\alpha_3$ ((a)-(d)), and System zero initial responses to probing input for $\alpha \in \mathcal{S}$ (e).}
    \label{Fig_probeResponse}
\vspace{-0em}
\end{figure}

 System responses to probing input and different random initial conditions under normal and different contingency scenarios are illustrated in Fig. \ref{Fig_probeResponse} (a) to (d), where the probing input response (red line) is dominant over the random initial condition responses (gray lines). Additionally, Fig.~\ref{Fig_probeResponse} (e) displays the zero initial probing input responses of $\alpha_0$ to $\alpha_3$ which exhibits the non-zero error of the system outputs under different contingencies. Note that, the system outputs illustrate the sum of local measurement deviations ($V_{DC}$, $I_{t,\{d,q\}}$, and $I_{L_{L4, \{d,q\}}}$) and system disturbances ($V_{a1,\{d,q\}}$). 

\begin{figure}[t]
    \vspace{-1em} % Reduce vertical space before the figure
    \centering
    \includegraphics[width=0.9\linewidth]{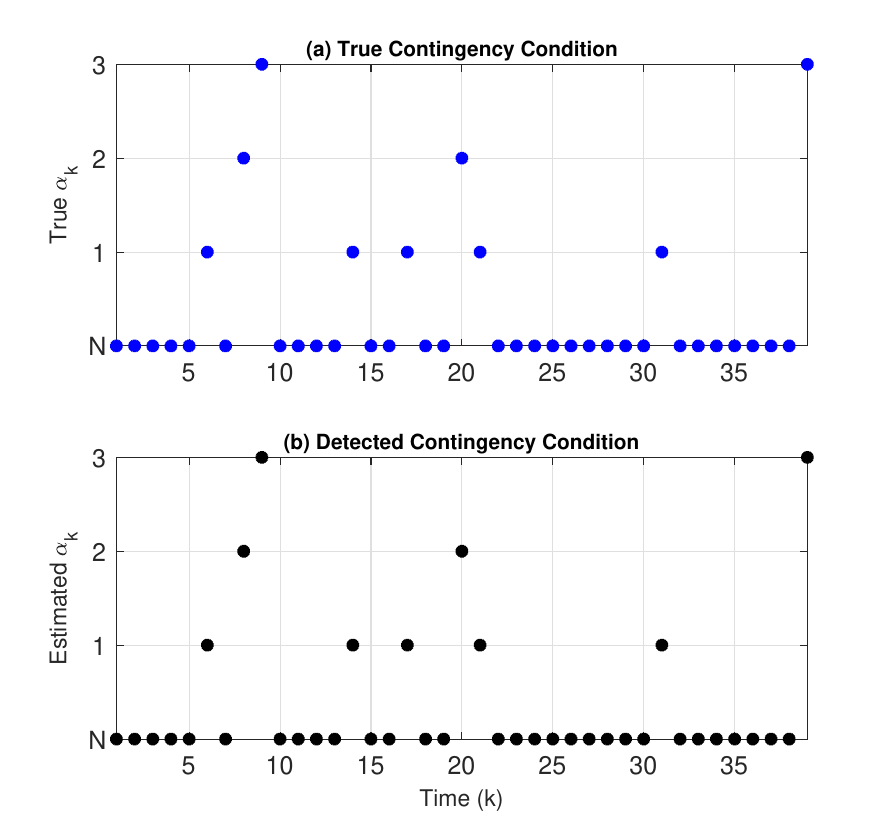}
    \vspace{-1.5em} % Reduce vertical space after the figure
    \caption{Power system random switching sequence between normal and contingency scenarios: (a) True Sequence, (b) Detected Sequence.}
    \label{Fig_Detection}
    \vspace{-1em}
\end{figure}

Finally, Fig. \ref{Fig_Detection}, illustrates the effectiveness of the contingency detection algorithm under MaMI probing input, showing the sequence of distribution system switching between different scenarios $\alpha \in \mathcal{S}$ every \(\tau = 0.6\) seconds, for time intervals \(k = 1 \ldots 40\). It can be observed that the contingency scenarios detected by implementing MaMI (black dots), can identify the true configurations of the system $\alpha_k$ (blue dots) without direct measurement of the contingency.

%~~~~~~~~~~~~~~~~~~~~~~~~~~~~~~~~~~
\section{Conclusion} \label{Sec5}

This paper presents a distributed approach for detecting unobservable contingencies in active distribution systems, leveraging limited measurements. Our approach is based on an SHS model that effectively captures various contingencies impact on the dynamics of PV-B resources, network connections, and loads.

This study emphasizes the necessity of MaMI probing inputs that effectively distinguish between contingencies and explores various probing types to identify the most effective approach for active distribution systems. The probing input enables the identification of the system's transfer function under each contingency. As each contingency corresponds to a distinct transfer function, the proposed probing methods ensure precise detection of contingencies that would otherwise remain undetected by conventional sensing and monitoring devices. We demonstrate the efficacy of the probing input and detection algorithm for a representative distribution system.

In the future, the SHS model will be utilized to enhance resilient power management in distributed systems under contingency scenarios. Advanced strategies will be employed to enable adaptive and robust control of the SHS framework.
\bibliographystyle{ieeetr}
\bibliography{references}
\end{document}